\documentclass[12pt]{article}

\usepackage{epsfig}
\usepackage{latexsym}
\usepackage{amsmath}
\usepackage{amssymb}
\usepackage{axodraw}

\providecommand{\ap}{{\alpha^\prime}}
\providecommand{\Dpa}{{\overline{Dp}}}
\providecommand{\Dta}{{\overline{D3}}}

\providecommand{\HF}{{\phantom{}^\star F}}

\providecommand{\HjDp}{{\phantom{}^\star j_{Dp}}}
\providecommand{\HjDdp}{{\phantom{}^\star j_{D(6-p)}}}

\providecommand{\kt}{{\kappa_{10}^2}}

\providecommand{\xDp}{{\vec{x}_\perp^{Dp}}}
\providecommand{\xDpz}{{\vec{x}_{\perp,0}^{Dp}}}

\providecommand{\detDp}{\frac{\sqrt{\det g_{Dp}}}{\sqrt{-\det g}}}

\providecommand{\tof}{{\sqrt{\frac{2}{5}}}}

\begin{document}

\begin{titlepage}
\begin{flushright}
\today
\end{flushright}

\vspace{1cm}
\begin{center}
\baselineskip25pt

{\Large\bf Schwarzschild Black Holes from Brane-Antibrane Pairs}

\end{center}
\vspace{1cm}
\begin{center}
\baselineskip12pt
{Axel Krause\footnote{E-mail: {\tt axkrause@x4u2.desy.de}}}
\vspace{1cm}

{\it Physics Department,}\\[1.8mm]
{\it National Technical University of Athens,}\\[1.8mm]
{\it 15773 Athens, Greece}

\vspace{0.3cm}
\end{center}
\vspace*{\fill}

\begin{abstract}
We show that D=4 Schwarzschild black holes can arise from a doublet of
Euclidean $D3$-$\Dta$ pairs embedded in D=10 Lorentzian spacetime.
By starting from a D=10 type IIB supergravity description for the
$D3-\Dta$ pairs and wrapping one of them over an external 2-sphere,
we derive all vacuum solutions compatible with the symmetry of the
problem. Analysing under what condition a Euclidean brane configuration
embedded in a Lorentzian spacetime can lead to a time-independent
spacetime, enables us to single out the embedded D=4 Schwarzschild
spacetime as the unique solution generated by the $D3$-$\Dta$ pairs. In
particular we argue on account of energy-conservation that
time-independent solutions arising from isolated Euclidean branes require
those branes to sit at event horizons. In combination with previous work
this self-dual brane-antibrane origin of the black hole allows for a
microscopic counting of its Bekenstein-Hawking entropy. Finally we
indicate how Hawking-radiation can be understood from the associated
tachyon condensation process.
\end{abstract}

\vspace*{\fill}

\end{titlepage}


\section{Introduction and Summary}
One of the most fascinating problems in fundamental physics today is to
find a proper set of microstates for spacetimes possessing an event
horizon such that the counting of their number matches the
Bekenstein-Hawking (BH) entropy
\begin{equation}
{\cal S}_{BH} = \frac{A_H}{4G_4} \; ,
\label{SBH}
\end{equation}
where $A_H$ is the area of the spacetime's event horizon. In the
framework of String-Theory such a set of entropy-carrying microscopic
states could be identified mostly for extremal black holes and
near-extremal ones (for some recent reviews see \cite{KS}). The basic
idea was to count microscopic states in the weakly coupled regime and to
export the result to the strongly coupled regime by using supersymmetry.
This made it difficult to employ the same strategy for highly non-extreme
spacetimes such as the Schwarzschild black hole or the de Sitter
universe. One of the approaches to deal with the Schwarzschild black
hole was to use certain boost-transformations and/or the T- and
S-dualities of String-Theory and thereby to achieve an entropy-preserving
mapping to a spacetime whose entropy-counting was under control, e.g.~the
three-dimensional BTZ black hole \cite{Sken} or near-extremal brane
configurations \cite{Arg}. During this process one however also had to
compactify time which might be justified from the Euclidean approach to
black hole thermodynamics. Other approaches within String-Theory to
microscopically derive the D=4 Schwarzschild BH-entropy had been
undertaken in \cite{Suss},\cite{Hal},\cite{SS},\cite{MAT},\cite{SS2}.

Finally, of course, one would like to know directly the microscopic
states in the strongly-coupled regime making up the black hole. Are they
located at or around the horizon like in the entanglement entropy
\cite{EEA},\cite{Muk}, thermal atmosphere \cite{TAA} or ``shape of the
horizon'' \cite{SHA} approaches? Or do they live elsewhere in the
interior or exterior region of a black hole? Unfortunately the answer to
this question is obscured by the indirect method of counting states
as has been pointed out in \cite{WLE}. Another puzzling aspect related to
the BH-entropy is its universality. One would like to know why it
universally applies not only to black holes but also to other spacetimes
possessing e.g.~cosmological event horizons.

In an approach to understand the origin and the universality of
the BH-entropy for spacetimes with spherical event horizons from
strongly coupled ($g_s\simeq 1$) String-/M-Theory, it was proposed in
\cite{K1} to consider doublets of orthogonal Euclidean dual brane pairs.
One brane of each pair had to wrap a sphere $S^2$ situated in the D=4
external spacetime while the complete 6- resp.~7-dimensional (for String-
resp.~M-Theory) internal part of spacetime was wrapped by the remaining
portion of the brane plus its dual brane partner. To connect this brane
picture to a D=4 spacetime with spherical event horizon $S^2_H$ whose
associated BH-entropy we would like to derive at a microscopic level,
it was proposed the following. The brane configuration would act as a
supergravity source capable of producing the D=4 spacetime in the
external part of its D=10 metric solution with $S^2\equiv S^2_H$
identified. Through the identification of a Euclidean brane's tension as
the inverse of a fundamental smallest volume unit it was then possible to
consider chain-states on the discretized branes' worldvolumes (see
\cite{Lind} for other indications of a discretized worldvolume at strong
coupling) and by counting their number to derive the D=4 BH-entropy plus
its logarithmic correction. This mechanism works for all D=4 spacetimes
as long as we can identify a doublet of dual Euclidean branes with the
D=4 spacetime under investigation in the prescribed manner. The main
purpose of this paper is to provide such an identification for the
case of the D=4 Schwarzschild black hole.

The organisation of the paper is as follows. In section 2 we will start
by analysing which branes and their duals might qualify for a potential
description of the D=4 Schwarzschild spacetime. Non-extremality and
charge-lessness of the latter leads us to consider an equal amount of
branes as antibranes while the property of being a vacuum solution (in
the Einstein equations sense) brings us to the non-dilatonic $D3$-$\Dta$
pairs. We then describe the D=10 geometry as appropriate for an exterior
solution with the branes-antibranes acting as the gravitational source.
In section 3 we derive all D=10 vacuum solutions which respect the
imposed symmetry -- among them a simple embedding of the D=4
Schwarzschild spacetime into D=10 spacetime. The Euclidean nature of our
considered branes and antibranes embedded in a D=10 Lorentzian spacetime
implies however a further constraint which we will discuss in section 5.
There we will argue that {\em an isolated Euclidean brane (or a finite
number of them) in a Lorentzian embedding violates energy-conservation
and leads to a time-dependence unless it is placed at an event horizon}.
On account of an infinite time-dilatation at the horizon, energy is
conserved from an outside observers point of view. At the same time the
ensuing geometry becomes time-independent. Demanding the existence of an
event horizon then singles out uniquely the embedded Schwarzschild
solution as representing the exterior geometry of our $D3$-$\Dta$
doublet. Therefore we might now invoke the results of \cite{K1} to derive
the BH-entropy of the Schwarzschild black hole and its logarithmic
corrections from counting chain-states on the branes-antibranes'
worldvolume. We end with section 6 by speculating on the origin of the
Hawking-radiation from the tachyon condensation point of view related to
the $D3$-$\Dta$ doublet.

\section{The Brane-Antibrane Pair Configuration}
In the following we will assume a type II String-Theory on a Lorentzian
spacetime manifold ${\cal M}^{1,3}\times{\cal M}^6$ where the internal
 space
${\cal M}^6$ is taken to be compact.

The D=4 Schwarzschild spacetime describes a non-rotating black hole which
bears no charges under any long-range gauge-field and is uniquely
characterized by its mass $M$. Moreover, it breaks all supersymmetry if
included as a background in String-Theory. In order to obtain this
spacetime from a doublet of dual brane pairs (the motivation for
this comes from the fact that such brane configurations allow for a
determination of the BH-entropy of the associated D=4 spacetime without
the need for supersymmetry \cite{K1}) of type II String-Theory, natural
candidates are the non-supersymmetric Euclidean brane-antibrane
configurations\footnote{See \cite{GM} for supergravity solutions
which interpolate between Lorentzian brane-antibrane pairs and
Schwarzschild black holes.}
\begin{equation}
(Dp,D(6-p))+(\overline{Dp},\overline{D(6-p)})
\end{equation}
or the fundamental string - NS5-brane configurations
\begin{equation}
(F1,NS5)+(\overline{F1},\overline{NS5}) \; , \qquad\qquad
(NS5,F1)+(\overline{NS5},\overline{F1}) \; ,
\label{NSSetup}
\end{equation}
where the first and second component in brackets are mutually
orthogonal. Moreover it is understood that the first component in each
bracket wraps an external $S^2$ contained in ${\cal M}^{1,3}$ while the
remaining worldvolume coordinates of both components cover the internal
${\cal M}^6$ completely \cite{K1}. Because we will drop the $F1$-$NS5$
pairs shortly, let us concentrate on the $Dp$-branes subsequently.

Setting all background tensor fields and the world-volume gauge-field
strength to zero except for the RR $(p+1)$-form $C_{p+1}$, the Euclidean
$Dp$-brane action in Einstein-frame is
\begin{equation}
S_{Dp}=T_{Dp}\int d^{p+1}xe^{(\frac{p-3}{4})\Phi}
\sqrt{\det g_{Dp}}+\mu_{Dp}\int C_{p+1} \; .
\end{equation}
with
\begin{equation}
T_{Dp}=\mu_{Dp}=\frac{1}{(2\pi)^p\ap^{(\frac{p+1}{2})}}
\end{equation}
the tension and charge of the brane, $\Phi$ the dilaton and $g_{Dp}$ the
induced metric on the brane. For the Euclidean antibrane $\overline{Dp}$
the sign of the second term is replaced by a minus. These specify the
sources of our system. To keep things simple let us start with a single
$Dp$-brane and add its dual plus antibrane partners step by step.

Together with the D=10 Einstein-frame supergravity bulk action for this
background (with a vanishing NS-NS 2-form potential $B$, in particular
the Chern-Simons terms of type II supergravity will be absent)
\begin{equation}
S_{SG}=\frac{1}{2\kt}\int \Big( eR
-\frac{1}{2}d\Phi\wedge\phantom{}^\star d\Phi
-\frac{1}{2}e^{a\Phi}F\wedge\HF \Big) \; ,
\label{Sugra}
\end{equation}
where $F=dC_{p+1}$, $e=\sqrt{-\det g}$ and $a$ is a constant depending on
the rank of $C_{p+1}$, this results in the following coupled Einstein-,
Maxwell- and dilaton equations
\begin{alignat}{3}
&R_{AB}-\frac{1}{2}Rg_{AB} =
\frac{e^{a\Phi}}{2(p+2)!} \Big( (p+2)F_{AA_2\hdots
A_{p+2}}{F_B}^{A_2\hdots A_{p+2}} -\frac{1}{2}g_{AB}F^2 \Big)
\notag \\
&+ \kt T_{Dp}e^{(\frac{p-3}{4})\Phi}\delta^{9-p}(\xDp-\xDpz)
\detDp\, g_{ij} \delta^i_A\delta^j_B
+ \frac{1}{2}\big( \partial_A\Phi\partial_B\Phi
-\frac{1}{2}g_{AB}\partial_C\Phi\partial^C\Phi \big) \; ,
\\
&e^{a\Phi}d\,\HF = 2\kt(-1)^p\,\HjDp \; ,
\\
&\Box\Phi = \frac{a}{2(p+2)!}e^{a\Phi} F^2 +
2\kt\delta^{9-p}(\xDp-\xDpz)T_{Dp}e^{(\frac{p-3}{4})\Phi}
\big(\frac{3-p}{4}\big)\detDp \; ,
\label{EOM}
\end{alignat}
where $i,j$ are indices along the brane, $A,B,\hdots$ are D=10
bulk indices, $\xDp$ denote all coordinates transverse to the
brane and $\xDpz$ give the brane localisation. Furthermore, $F^2\equiv
F_{A_1\hdots A_{p+2}}F^{A_1\hdots A_{p+2}}$ and the brane-current
$j_{Dp}$ is given by
\begin{equation}
j_{Dp}=\mu_{Dp}\delta^{10-(p+1)}(\xDp-\xDpz)\detDp\omega_{Dp}
\end{equation}
with $\omega_{Dp}=\sqrt{\det g_{Dp}}dx^{i_1}\wedge\hdots\wedge
dx^{i_{p+1}}$ the positively oriented metric volume element on the
$Dp$-brane's worldvolume.

If we now add an anti-$\Dpa$-brane which coincides with the $Dp$-brane
then due to the opposite RR-charges $F$ vanishes. This is consistent with
the Bianchi-identity which receives a magnetic source term from the dual
$D(6-p)$-brane
\begin{equation}
e^{a\Phi}dF=-2\kt\HjDdp
\end{equation}
but which also gets compensated by adding the
anti-$\overline{D(6-p)}$-brane. Thus we can neglect the field-strength
$F$ in the above field equations (and similarly the hitherto
suppressed dual field-strength associated with the dual brane $D(6-p)$).
Moreover, we are interested in the solution exterior to the sources,
i.e.~for $\xDp\ne\xDpz$, describing the long-range fields. Hence we can
also drop the brane source terms which means that we are left with
\begin{alignat}{3}
&R_{AB}=\frac{1}{2}\partial_A\Phi\partial_B\Phi \\
&\Box\Phi =0 \; .
\end{alignat}
From the Einstein equations we recognize that we have to demand a
constant dilaton in order to obtain a vacuum solution like Schwarzschild
(there might also be dilatonic brane-antibrane configurations with
$\Phi=\Phi(x_4,\hdots,x_9)$ such that $R_{\mu\nu}=0$ and $R_{mn}\ne 0$
($\mu,\nu=0,\hdots,3;\; m,n=4,\hdots,9$) but these will not be studied
here). Without loss of generality we can then set $\Phi=0$ and end up
with the D=10 vacuum Einstein equations
\begin{equation}
R_{AB}=0\; .
\end{equation}

It is well-known that the only non-dilatonic $Dp$-brane is the self-dual
$D3$-brane. Thus our dual brane pair doublet will consist of two
$D3$-$\Dta$ Euclidean brane-antibrane pairs located at some finite value
of $r$, the D=4 radial distance, as depicted in fig.\ref{Setup}. This
non-dilatonic property also excludes the second possibility
(\ref{NSSetup}).
\setcounter{figure}{0}
\begin{figure}[t]
\begin{center}
\begin{picture}(215,120)(0,0)
\Text(30,72)[]{$t$}
\Text(50,72)[]{$r$}
\Text(70,72)[]{$\theta$}
\Text(90,72)[]{$\phi$}
\Text(110,72)[]{$4$}
\Text(130,72)[]{$5$}
\Text(150,72)[]{$6$}
\Text(170,72)[]{$7$}
\Text(190,72)[]{$8$}
\Text(210,72)[]{$9$}

\Text(2,54)[]{$D3$}
\Text(2,36)[]{$\Dta$}
\Text(2,18)[]{$D3$}
\Text(2,0)[]{$\Dta$}

\Line(-3,63)(215,63)
\Line(17,-6)(17,74)

\Text(70,54)[]{$\bullet$}
\Text(90,54)[]{$\bullet$}
\Text(110,54)[]{$\bullet$}
\Text(130,54)[]{$\bullet$}

\Text(70,36)[]{$\bullet$}
\Text(90,36)[]{$\bullet$}
\Text(110,36)[]{$\bullet$}
\Text(130,36)[]{$\bullet$}

\Text(150,18)[]{$\bullet$}
\Text(170,18)[]{$\bullet$}
\Text(190,18)[]{$\bullet$}
\Text(210,18)[]{$\bullet$}

\Text(150,0)[]{$\bullet$}
\Text(170,0)[]{$\bullet$}
\Text(190,0)[]{$\bullet$}
\Text(210,0)[]{$\bullet$}
\end{picture}
\caption{The two Euclidean $D3$-$\Dta$ brane-antibrane pairs are oriented
along the directions marked by dots. The coordinates $t,r,\theta,\phi$
describe the D=4 portion with $\theta,\phi$ describing the $S^2$. The
whole configuration is located at some common fixed value of $r$.}
\label{Setup}
\end{center}
\end{figure}
One might think that a brane-antibrane pair would annihilate itself
within a very short lifetime. How this lifetime gets infinitely
extended will be addressed later on in section 4. It might also be of
interest to study the effect on the geometry when more background fluxes
are switched on (see e.g.~\cite{KT}) and lower-dimensional branes are
induced as suggested by K-Theory \cite{WittK}. However, this is beyond
the scope of the present paper.

In order to solve the vacuum equations in the exterior region of the
sources, let us make an Ansatz for the D=10 geometry reflecting the
symmetry-properties of the two Euclidean $D3$-$\Dta$ pairs. Searching for
a time-independent static solution (actually time-independence in the
context of Euclidean branes poses a further constraint on the location
of the brane to be imposed later on) and noting that the only spatial
coordinate transverse to all branes is the D=4 radial coordinate $r$, the
Ansatz should be
\begin{equation}
ds^2=
-e^{2\lambda(r)}dt^2+e^{2\nu(r)}dr^2+\big(r^2d\Omega^2+dx_4^2
+dx_5^2\big) +e^{2\psi(r)}\big( dx_6^2+\hdots +dx_9^2 \big) \; ,
\label{Metric}
\end{equation}
with $d\Omega^2=d\theta^2+\sin^2\theta d\phi^2$ the metric of a unit
two-sphere. The fact that no further $r$-dependent factor multiplies the
$\theta,\phi,x_4,x_5$ worldvolume of the one $D3$-$\Dta$ pair is because
such a factor can be set equal to one by a redefinition of $r$ and
further noticing that the $D3$-$\Dta$ worldvolume has to be multiplied
as a whole by a common factor.

\section{The Vacuum Solutions}
The D=10 Einstein vacuum equations for this metric deliver the following
second order ordinary differential equations (ODE's)
\begin{alignat}{3}
&\Lambda'+\Lambda^2-\Lambda N+2\frac{\Lambda}{r}+4\Lambda\Psi = 0
\label{t}\\
&\Lambda'+\Lambda^2-\Lambda N-2\frac{N}{r}+4\Psi'+4\Psi^2-4N\Psi = 0
\label{r}\\
&\Lambda-N+\frac{1}{r}(1-e^{2\nu})+4\Psi = 0
\label{theta}\\
&\Lambda\Psi+\Psi'+4\Psi^2-N\Psi+2\frac{\Psi}{r} = 0
\label{6}
\end{alignat}
where we have defined for convenience
\begin{equation}
\Lambda = \lambda', \quad N = \nu', \quad \Psi = \psi' \; .
\end{equation}

Together with these ODE's we have to choose an appropriate boundary
condition at some value of $r$. We should expect that if we depart
sufficiently far from the brane configuration towards
larger $r$ that the influence of the gravitational source becomes smaller
and smaller until finally the D=10 spacetime should approach flat
Minkowski spacetime. Thus our boundary conditions will be asymptotic
flatness at $r\rightarrow\infty$
\begin{equation}
\lambda\rightarrow 0 \; , \;\;
\nu\rightarrow 0 \; , \;\;
\psi\rightarrow 0 \; .
\label{ABC}
\end{equation}

Our task in the following will be first to find all solutions to this
set of vacuum equations which are compatible with asymptotic flatness and
then second to select that subclass of solutions which actually qualifies
as an exterior solution sourced by the two Euclidean $D3-\Dta$ pairs by
examining under what condition Euclidean branes can lead to a stationary
solution in coordinates adapted to an asymptotic observer.

To start with, one derives from the linear combination
$(\ref{t})-(\ref{theta})\times\Lambda$ that
\begin{equation}
\Lambda'+\frac{\Lambda}{r}(1+e^{2\nu})=0
\end{equation}
while the combination $(\ref{6})-(\ref{theta})\times\Psi$ leads to
\begin{equation}
\Psi'+\frac{\Psi}{r}(1+e^{2\nu})=0 \; .
\label{TWO}
\end{equation}
From the formal solution of these two equations for $\Lambda$ and $\Psi$
one deduces that we face four different cases. Either $\Lambda$ or $\Psi$
or both are equal to zero or otherwise they have to be proportional to
each other. We will now study these different cases in detail.

\subsection{The Case $\Lambda=\Psi=0$}
Let us begin with the simplest case of $\Psi=\Lambda=0$. It is easy to
see that here the Einstein vacuum equations produce just the D=10 flat
Minkowski metric
\begin{equation}
ds^2 = -dt^2+dr^2+r^2d\Omega^2+dx_4^2+dx_5^2+dx_6^2+\hdots +dx_9^2 \; .
\end{equation}
with spacetime topology $M^{1,3}\times T^6$ due to compactness of
the internal space.

\subsection{The Case $\Lambda\ne 0,\,\Psi=0$}
This is the next easiest case which will bring us to the D=4
Schwarzschild solution. Because $\Psi=0$, $\psi$ will be constant and
can be set to zero by an appropriate scaling of the $x_6,\hdots,x_9$
coordinates. By subtracting $(\ref{t})-(\ref{r})$, we obtain $\Lambda=-N$
which amounts to the relation
\begin{equation}
\lambda =-\nu
\end{equation}
as an additive integration constant again can be absorbed into a
redefinition of the coordinates. With this input, (\ref{theta}) leads to
the ODE for $\nu$
\begin{equation}
\nu' = \frac{1}{2r}(1-e^{2\nu})
\end{equation}
which is solved by
\begin{equation}
e^{2\nu} = \big(1\pm \frac{r_0}{r}\big)^{-1}
\end{equation}
with a positive $r_0$. Altogether, we end up with an embedding (``lift'')
of the D=4 Schwarzschild solution into D=10 spacetime
\begin{equation}
ds^2 =
-\big(1\pm\frac{r_0}{r}\big)dt^2+\big(1\pm\frac{r_0}{r}\big)^{-1}dr^2
+r^2d\Omega^2+dx_4^2+dx_5^2 +dx_6^2+\hdots +dx_9^2 \; .
\label{SS}
\end{equation}
Again the internal six-dimensional space is topologically $T^6$. The
solution with the plus-sign would correspond (if one could smooth out its
naked singularity) to a source with negative mass. It cannot describe
the $D3$-$\Dta$ geometry as all (anti-)branes involve only positive
tensions and will therefore be ruled out.

An alternative characterisation of this spacetime is as a black D6-brane
in its ultra non-extreme limit. In this limit the black D6-brane looses
its magnetic Ramond-Ramond 2-form charge while the dilaton becomes
constant thus giving a non-dilatonic vacuum solution.

\subsection{The Case $\Lambda=0,\,\Psi\ne 0$}
In this case $\lambda$ has to be constant and without loss of generality
can be set to $\lambda = 0$. It therefore remains to determine $\nu$ and
$\psi$.

Of the vacuum Einstein equations, (\ref{t}) is identically satisfied. The
remaining three equations can be chosen conveniently as
$(\ref{r})-4\times(\ref{6})$, (\ref{theta}) and $(\ref{6})/\Psi$
\begin{alignat}{3}
&N=-2\Psi(3\Psi r+2)
\label{a} \\
&N=4\Psi+\frac{1}{r}(1-e^{2\nu})
\label{b} \\
&N=(\ln(r_1\Psi))'+4\Psi+\frac{2}{r}
\label{g} \; .
\end{alignat}
Here a positive constant $r_1$ with dimension of a length has been
introduced to keep the function inside the logarithm dimensionless.
Clearly we have an overconstrained system with three equations in two
unknowns. The last equation can be integrated directly to yield
after exponentiation
\begin{equation}
(e^{4\psi})'=4\,\frac{r_1}{r^2}\,e^\nu
\label{g2} \; ,
\end{equation}
where the exponentiated integration constant can be identified with
$r_1$. From this equation it is easy to see that $\Psi\ge 0$.
Consequently (\ref{a}) demands that $N\le 0$ and therefore from (\ref{b})
it follows that
\begin{equation}
e^{2\nu}\ge 1 \; .
\label{INE}
\end{equation}

Unfortunately eliminating $N$ in the above equations and solving for
$\Psi$ leads to rather complicated differential equations. Let us
therefore proceed by using the first two equations to eliminate $\Psi$
and obtain a differential equation for $\nu$. More specifically we solve
(\ref{b}) for $\Psi$ and with this result eliminate $\Psi$ in (\ref{a}).
The ensuing quadratic equation in $N=\nu'$ possesses the solution
\begin{equation}
-\frac{3}{8}\nu' r =H(\nu)(H(\nu)\pm 1) \; ,
\label{ODE4}
\end{equation}
where
\begin{equation}
H(\nu)\equiv\sqrt{5+3e^{2\nu}}/\sqrt{8} \; .
\label{Gdef1}
\end{equation}
Notice that $H\ge 1$ due to (\ref{INE}).
In order to resolve the sign ambiguity let us consider its asymptotics.
From the asymptotic boundary conditions (\ref{ABC}) we infer that
$H\rightarrow 1$. Thus we obtain from (\ref{ODE4}) for $\nu$ the
asymptotic behaviour $\nu\simeq -\frac{8}{3}(1\pm 1)\ln r$ which only
gives $\nu\rightarrow 0$ if we choose the minus-sign in
(\ref{ODE4}).

Employing the chain-rule the above ODE for $\nu$ translates into an ODE
for $H(r)\equiv H(\nu(r))$
\begin{equation}
\frac{H'(r)}{(H(r)-1)(5-8H^2(r))}=\frac{1}{3r} \; .
\label{GODE1}
\end{equation}
By standard integration techniques this ODE can be
shown to possess the following implicit solution
\begin{equation}
\frac{1}{(H(r)-1)}\frac{(H(r)-\sqrt{\frac{5}{8}})^{\tof +\frac{1}{2}}}
                       {(H(r)+\sqrt{\frac{5}{8}})^{\tof -\frac{1}{2}}}
=\frac{r}{r_2} \; .
\label{Sol2}
\end{equation}
with positive $r_2$. By rewriting (\ref{Gdef1}) as
\begin{equation}
e^{2\nu}=\frac{1}{3}(8H^2(r)-5)
\label{Gdef2}
\end{equation}
we gain the solution for the radial component of the metric.

Next we have to determine $\psi$ which we will do by starting from
(\ref{g2}). Using (\ref{Gdef2}), (\ref{GODE1}) and applying the
chain-rule to rewrite the derivative as a derivative with respect to $H$,
we obtain
\begin{equation}
\frac{d(e^{4\psi})}{dH} = -\frac{12r_1}{\sqrt{3}r(H-1)\sqrt{8H^2-5}}
\; .
\label{GODE}
\end{equation}
The $r$ dependence which hinders a straightforward integration can be
transformed into a $H$ dependence by means of the solution (\ref{Sol2}),
giving the ODE
\begin{equation}
\frac{d(e^{4\psi})}{dH}=-\frac{\sqrt{6}r_1}{r_2}
\frac{(H+\sqrt{\frac{5}{8}})^{\tof -1}}{(H-\sqrt{\frac{5}{8}})^{\tof +1}}
\; .
\end{equation}
By an auxiliary transformation,
$H=\sqrt{\frac{5}{2}}(\frac{1}{(x-1)}+\frac{1}{2})$, of the integration
variable $H$ to a new variable $x$, this ODE can trivially be integrated
with the result
\begin{equation}
e^{2\psi} = \bigg( \sqrt{6}\frac{r_1}{r_2}
\Big[ h(1)-h(H(r)) \Big] + 1 \bigg)^{\frac{1}{2}} \; , \qquad
h(y)\equiv \Big(\frac{\tof y +\frac{1}{2}}{\tof y
-\frac{1}{2}}\Big)^{\tof}  \; .
\end{equation}

However, we will now show that the overconstrained system of equations
(\ref{a}),(\ref{b}),(\ref{g}) requires a much more stringent condition on
$H(r)$. Namely with help of (\ref{GODE1}) it turns out that (\ref{TWO})
or equivalently the difference $(\ref{g})-(\ref{b})$ amounts to
\begin{equation}
(H(r)-1)\bigg(\frac{r_2+\sqrt{6}r_1h(1)}{r_2e^{4\psi}}
\bigg) = 0 \; ,
\end{equation}
which requires setting $H(r)\equiv 1$. This would render the metric
trivial and therefore violates our assumption that $\Psi\ne 0$. Hence we
can conclude that there is no solution for the case $\Lambda =0,\Psi\ne
0$.

\subsection{The Case $\Lambda\ne 0,\,\Psi\ne
 0,\;\Psi=c\Lambda,\,c\ne -\frac{1}{4},-\frac{2}{3},0$}
Due to the proportionality of $\Psi$ and $\Lambda$ the D=10 Einstein
vacuum equations reduce again to a system of three equations in two
unknowns
\begin{alignat}{3}
&\Lambda'+a\Lambda^2+\Lambda\big(\frac{2}{r}-N\big) = 0
\label{t2}\\
&a\Lambda'+(1+4c^2)\Lambda^2-a\Lambda N-\frac{2}{r}N = 0
\label{r2}\\
&\Lambda'+\frac{1}{r}(1+e^{2\nu})\Lambda = 0 \; .
\label{c2}
\end{alignat}
with the constant coefficients\footnote{Notice that the parameter $a$
defined here has nothing to do with the constant $a$ appearing in
(\ref{Sugra})--(\ref{EOM}).}
\begin{equation}
a=1+4c \; , \qquad b=4c(2+3c) \; .
\end{equation}
The combination $2a^2-b>0$ which will appear below is
positive for any value of $c$. Moreover, $c\ne 0$ implies $a\ne 0$ and
$b\ne 0$ if we exclude the case $c=-\frac{2}{3}$ which will be
treated separately.

These equations can be brought into a more manageable form by taking the
linear combinations $a\times (\ref{t2})-(\ref{r2})$ and
$((\ref{t2})-(\ref{c2}))/\Lambda$
\begin{alignat}{3}
&b\Lambda^2+\frac{2}{r}(a\Lambda+N) = 0
\label{t3}\\
&a\Lambda = \frac{1}{r}(e^{2\nu}-1)+N
\label{r3}\\
&\Lambda'+\frac{1}{r}(1+e^{2\nu})\Lambda = 0 \; .
\label{c3}
\end{alignat}
Now we can multiply the first equation by $a^2$ and eliminate $a\Lambda$
in it by substituting the second equation. Some minor manipulations then
bring (\ref{t3}) into the following ODE for $\nu$
\begin{equation}
(br\nu'+k(\nu))^2 = 2a^2k(\nu) \; ,
\end{equation}
where
\begin{equation}
k(\nu) \equiv 2a^2+b(e^{2\nu}-1) \; .
\end{equation}
Due to the square on the lhs also the rhs has to be non-negative.
Thus it proves convenient to define a new non-negative function
\begin{equation}
K(\nu) = \sqrt{k(\nu)}
\end{equation}
such that the ODE for $\nu$ becomes
\begin{equation}
br\nu' = -K(\nu)(K(\nu)\pm\sqrt{2}|a|) \; .
\label{ODE1}
\end{equation}
The sign-ambiguity in this equation which results from taking the
square-root can be resolved once more by examining its asymptotic
behaviour. At $r\rightarrow\infty$ the asymptotic boundary condition
demands $\nu\rightarrow 0$ and thus that $K(\nu)$ approaches
\begin{equation}
K(\nu)\simeq\sqrt{2}|a| \; .
\end{equation}
Hence we can deduce from (\ref{ODE1}) that in the asymptotic regime
$b\nu\simeq 2a^2(1\pm 1)\ln r$. To satisfy the asymptotic boundary
condition for $\nu$ we therefore have to choose the minus sign in
(\ref{ODE1}). We can now rewrite (\ref{ODE1}) as an ODE for $K(r)\equiv
K(\nu (r))$ by employing the chain-rule
\begin{equation}
\frac{K'(r)}{(\sqrt{2}|a|-K(r))(K^2(r)-2a^2+b)}=\frac{1}{br} \; .
\label{KODE}
\end{equation}
By standard integration techniques this leads to the following implicit
solution for $K(r)$
\begin{equation}
\frac{1}{|\sqrt{2}|a|-K(r)|}
\frac{|\sqrt{2a^2-b}-K(r)|^{\frac{|a|}{\sqrt{2(2a^2-b)}}+\frac{1}{2}}}
     {(\sqrt{2a^2-b}+K(r))^{\frac{|a|}{\sqrt{2(2a^2-b)}}-\frac{1}{2}}}
=\frac{r}{r_3}
\label{Sol1}
\end{equation}
with an integration constant $r_3>0$. Knowing $K(r)$ the radial
component of the metric is then obtained simply by rewriting the defining
equation for $K(r)$ as
\begin{equation}
e^{2\nu}=\frac{K^2(r)-2a^2+b}{b} \; .
\label{Nu}
\end{equation}

It remains to determine $\lambda$ and $\psi$. For the former we use
(\ref{r3}) and substitute $e^{2\nu}$ in it through (\ref{Nu}). This
gives
\begin{equation}
a\lambda'=\frac{1}{r}\big(\frac{K^2-2a^2}{b}\big)+\nu' \; .
\label{H1}
\end{equation}
We will now bring the term in brackets into a form which can be easily
integrated. To this aim we use (\ref{KODE}) which allows us to bring
(\ref{H1}) into the form
\begin{equation}
a\lambda'=\frac{K'}{\sqrt{2}|a|-K}-\frac{1}{r}+\nu'
\end{equation}
which can be integrated directly to give
\begin{equation}
\lambda =-\frac{1}{a}\ln|(\sqrt{2}|a|-K)\frac{r}{r_4}|+\frac{\nu}{a}
\end{equation}
with an integration constant $r_4>0$. Once more using (\ref{Nu}) to
express $\nu$ in terms of $K$, we obtain the time-component of the metric
\begin{equation}
e^{2\lambda}=\Big(\big(\frac{r_4}{r}\big)^2
\frac{K^2(r)-2a^2+b}{b(K(r)-\sqrt{2}|a|)^2}\Big)^{\frac{1}{a}} \; .
\end{equation}

Finally, since we have $\Psi=c\Lambda$, the internal metric component
$e^{2\psi}$ is related to the time-component via
\begin{equation}
e^{2\psi}=(e^{2\lambda})^c
\end{equation}
where we used the freedom of scaling coordinates to drop the
integration constant.

It is not hard to check that the results obtained for
$e^{2\lambda},e^{2\nu},e^{2\psi}$ indeed satisfy all three equations of
the overconstrained system (\ref{t3}),(\ref{r3}),(\ref{c3}). Consequently
we have the following class of vacuum solutions in this case
\begin{alignat}{3}
ds^2 = -\,&\Big(\big(\frac{r_4}{r}\big)^2\frac{K^2(r)-2a^2+b}
{b(K(r)-\sqrt{2}|a|)^2}\Big)^{\frac{1}{a}}dt^2
+\Big(\frac{K^2(r)-2a^2+b}{b}\Big)dr^2
+(r^2d\Omega^2+dx_4^2+dx_5^2) \notag \\
+&\,\Big(\big(\frac{r_4}{r}\big)^2\frac{K^2(r)-2a^2+b}
{b(K(r)-\sqrt{2}|a|)^2}\Big)^{\frac{c}{a}}(dx_6^2+\hdots+dx_9^2)
\label{MetricSol4}
\end{alignat}
provided that $a\ne 0,\,b\ne 0$.

We should add that in order to obtain an asymptotically flat vacuum
solution one has to relate the two positive constants $r_3$ and $r_4$
in the following way
\begin{equation}
r_4 = r_3
\frac{\big|\sqrt{2a^2-b}-\sqrt{2}|a|\big|^{\frac{|a|}{\sqrt{2(2a^2-b)}}
+\frac{1}{2}}}
{\big(\sqrt{2a^2-b}+\sqrt{2}|a|\big)^{\frac{|a|}{\sqrt{2(2a^2-b))}}
-\frac{1}{2}}}
\end{equation}
which can be seen from an explicit derivation of the weak-field limit of
(\ref{MetricSol4}).

\subsection{The Case $\Lambda\ne 0,\,\Psi\ne 0,\;\Psi=-\frac{1}{4}\Lambda$}
The case $c=-\frac{1}{4}$ corresponds to the situation where $a=0$ which
we had omitted previously. Here the vacuum Einstein equations reduce to
the equations
\begin{alignat}{3}
&\Lambda'+\Lambda\big(\frac{2}{r}-N\big) = 0
\label{t4}\\
&\frac{5}{4}\Lambda^2-\frac{2}{r}N = 0
\label{r4}\\
&\Lambda'+\frac{1}{r}(1+e^{2\nu})\Lambda = 0 \; .
\label{c4}
\end{alignat}
Subtracting the first from the third equation leads to
\begin{equation}
\nu'=\frac{1}{r}(1-e^{2\nu}) \; ,
\label{ODE3}
\end{equation}
which is solved by
\begin{equation}
e^{2\nu}=\big(1\pm(\frac{r_5}{r})^2\big)^{-1}
\label{SOL5}
\end{equation}
with $r_5$ a positive integration constant. Notice that the rhs of
(\ref{ODE3}) is twice as large as in the Schwarzschild case and
consequently leads to the quadratic dependence on $r$ instead of a linear
dependence as for Schwarzschild.

Next, we determine $\lambda$ from (\ref{r4}) by using the solution
for $\nu$
\begin{equation}
\lambda = \pm 2\sqrt{\frac{2}{5}}
\ln\Big(\frac{r}{r_5+\sqrt{r_5^2\pm r^2}}\Big) \; .
\end{equation}
Notice that the sign-ambiguity in front is unrelated to the one under the
square-root which coincides with the one of (\ref{SOL5}). By appealing
to a scaling of $t$, we have suppressed the integration constant. Thus we
obtain four different solutions. However those with a minus-sign under
the square-root have a restricted range $r\le r_5$ and can therefore (in
these coordinates) not reach spatial infinity. Because there is no reason
why without any further gravitational sources spacetime outside the
brane-antibrane configuration should abruptly come to an end, we will
discard these as exterior solutions.

Finally, we have $\psi=-\frac{1}{4}\lambda$, once more suppressing the
integration constant and therefore arrive at the following vacuum
solution (one can check that it satisfies the complete set of
overconstrained equations (\ref{t4}),(\ref{r4}),(\ref{c4}))
\begin{alignat}{3}
ds^2 = - &\Big(\frac{r}{r_5+\sqrt{r_5^2+r^2}}\Big)^{\pm 4\tof}dt^2
+\big(1+(\frac{r_5}{r})^2\big)^{-1}dr^2+(r^2d\Omega^2+dx_4^2+dx_5^2)
\notag \\
+ &\Big(\frac{r}{r_5+\sqrt{r_5^2+r^2}}\Big)^{\mp\tof}
(dx_6^2+\hdots +dx_9^2) \; .
\label{MetricSol5}
\end{alignat}

\subsection{The Case $\Lambda\ne 0,\,\Psi\ne 0,\;\Psi=-\frac{2}{3}\Lambda$}
This is the second case which we had left out before and it corresponds
to setting $b=0$. In this case the Einstein equations
(\ref{t3}),(\ref{r3}),(\ref{c3}) amount to
\begin{alignat}{3}
&\frac{5}{3}\Lambda = N
\label{t5}\\
-&\frac{5}{3}\Lambda = N+\frac{1}{r}(e^{2\nu}-1)
\label{r5}\\
&\Lambda'+\frac{1}{r}(1+e^{2\nu})\Lambda = 0 \; .
\label{c5}
\end{alignat}
The first equation leads straight to $\frac{5}{3}\lambda =\nu$ while
adding (\ref{t5}) and (\ref{r5}) gives
\begin{equation}
\nu' = \frac{1}{2r}(1-e^{2\nu}) \; .
\end{equation}
This is the same ODE as in the Schwarzschild case and gets solved by
\begin{equation}
e^{2\nu}=\big(1\pm\frac{r_6}{r}\big)^{-1} \; ,
\end{equation}
where $r_6>0$. With $\lambda =\frac{3}{5}\nu$ and $\psi
=-\frac{2}{3}\lambda$ one can check that also (\ref{c5}) is satisfied and
one finds the vacuum solution
\begin{alignat}{3}
ds^2 = - &\big(1\pm\frac{r_6}{r}\big)^{-\frac{3}{5}}dt^2
+\big(1\pm\frac{r_6}{r}\big)^{-1}dr^2+(r^2d\Omega^2+dx_4^2+dx_5^2)
\notag \\
+ &\big(1\pm\frac{r_6}{r}\big)^{\frac{2}{5}}(dx_6^2+\hdots +dx_9^2) \; .
\label{MetricSol6}
\end{alignat}

\section{Time-Dependence, Energy-Conservation and Euclidean Branes at the
Horizon}
In order to discriminate which one of the solutions obtained in
the previous section actually represents the exterior geometry of the
Euclidean $D3$-$\Dta$ pair doublet, we will now examine under what
condition such a doublet can lead to a stationary, i.e.~time-independent
geometry.

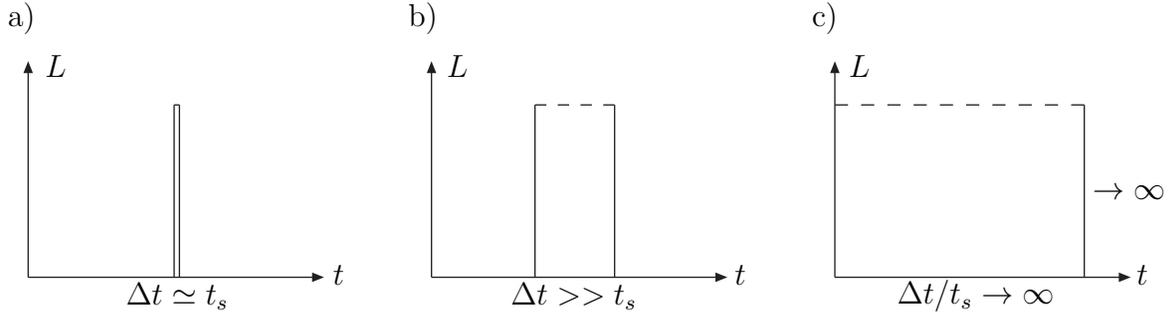
\begin{figure}[t]
\begin{center}
\begin{picture}(215,120)(0,0)

\LongArrow(-100,0)(-100,80)
\LongArrow(-100,0)(10,0)
\Text(-103,98)[]{a)}
\Text(-90,80)[]{$L$}
\Text(17,1)[]{$t$}
\Line(-45,0)(-45,65)
\Line(-43,0)(-43,65)
\DashLine(-45,65)(-43,65){5}
\Text(-44,-7)[]{$\Delta t \simeq t_s$}

\LongArrow(52,0)(52,80)
\LongArrow(52,0)(162,0)
\Text(49,98)[]{b)}
\Text(62,80)[]{$L$}
\Text(169,1)[]{$t$}
\Line(121,0)(121,65)
\Line(91,0)(91,65)
\DashLine(91,65)(121,65){5}
\Text(106,-7)[]{$\Delta t >> t_s$}

\LongArrow(204,0)(204,80)
\LongArrow(204,0)(314,0)
\Text(201,98)[]{c)}
\Text(214,80)[]{$L$}
\Text(321,1)[]{$t$}
\Line(298,0)(298,65)
\Text(316,32)[]{$\rightarrow\infty$}
\DashLine(204,65)(298,65){5}
\Text(258,-7)[]{$\Delta t/t_s\rightarrow\infty$}
\end{picture}
\caption{a) An isolated Euclidean brane at weak coupling where gravity is
nearly switched off. Its lifetime $\Delta t$ is of the order of the
string-time, $t_s=\sqrt{\ap}$, which violates energy-conservation. b) A
Euclidean brane in the presence of a strong gravitational field. Due to
time dilatation its lifetime increases. c) A Euclidean brane sitting at
an event horizon. Its lifetime becomes infinite and the configuration
becomes stationary. Energy-conservation gets restored. The ordinate $L$
stands for all the longitudinal space directions of the Euclidean brane
while the abscissa gives an asymptotic observer's time $t$.}
\label{EB}
\end{center}
\end{figure}

In as much as a conventional D-brane breaks translational symmetry
orthogonal to its worldvolume a Euclidean D-brane embedded in a
Lorentzian spacetime breaks furthermore time-translation symmetry because
time is now a transverse coordinate. A breaking of space-translation
symmetry results in a violation of momentum conservation for those
momentum components orthogonal to the D-brane's worldvolume. Hence, in
the case of a Euclidean D-brane also energy-conservation would be
violated. This is most obvious in the weakly coupled ($g_s\ll 1$)
regime where a Euclidean D-brane satisfies a Dirichlet-boundary condition
in time, i.e.~it exists only for a snapshot at a moment in time --
before or after it is non-existent. This sudden creation out of nothing
and instantaneous destruction afterwards clearly violates
energy-conservation (see fig.\ref{EB}a).

In the case of a Euclidean brane-antibrane pair in flat spacetime there
are two notions of lifetime. One is related to its Euclidean nature as
explained before, the other is related to its decay via tachyon
condensation (see \cite{Strom} for the latter). Presumably both notions
coincide and are of order the string-time $t_s=\sqrt{\ap}$. For the
brane-antibrane pair energy is conserved when it decays into radiation,
however energy-conservation becomes a problem when the pair becomes
created out of nothing due to its Euclidean nature. This might be cured
by some finely tuned incoming radiation \cite{Strom} but there may be
doubts whether the entropy in this process ``radiation $\rightarrow$
brane-antibrane pair'' is decreasing thus violating the second law of
thermodynamics.

In the presence of gravitational fields we are used to the phenomenon of
time-dilatation. Accordingly the lifetime of a Euclidean brane could be
enhanced (from the point of view of an exterior observer) in a strong
gravitational field (see fig.\ref{EB}b). Though the two incidents of
energy-conservation violation (the creation and later annihilation of the
brane) become more separated in an exterior observer's time it is still
present.

There is however one unique possibility of obtaining a consistent theory
with Euclidean branes and at the same time obtaining a stationary
geometry. This is when the Euclidean branes are located precisely at a
spacetime event horizon (see fig.\ref{EB}c). At an event horizon
the gravitational redshift becomes infinitely large thus rendering a
Euclidean brane's lifetime infinitely large. The troublesome
energy-violating incidences become removed to $t\rightarrow\pm\infty$
which means they do not occur. From an exterior (to the brane) observer's
point of view time on the Euclidean brane stands still and so the
configuration becomes time-independent, i.e.~stationary. In the same
manner the short lifetime of a brane-antibrane due to annihilation gets
infinitely enhanced from the an exterior observer's view and thus
its exterior geometry becomes actually time-independent at a classical
level.

Due to this reasoning which suggests that {\em isolated Euclidean branes
embedded in a Lorentzian spacetime should sit at event horizons} we
will impose on our vacuum solution the requirement that it should possess
such an event horizon where we can locate the Euclidean
brane-antibrane pairs (notice that all our vacuum solutions are given in
asymptotically flat coordinates as appropriate for an external observer).
Due to the location of the $D3-\Dta$ pairs, the horizon must appear at
some finite value of $r$. As an aside we want to remark that one could
also think of an array of Euclidean branes along the time direction with
the effect to arrive at a smeared out brane with no time-dependence any
more and thus also a stationary geometry. This however requires an
infinity of Euclidean branes which is not what we have here.

In order to examine which of our solutions exhibits the required
event horizon let us mention two necessary criteria for such a
horizon. Consider radial null curves for which $ds^2=0$ and all
coordinates are constant except for $t$ and $r$. This leads to
\begin{equation}
\frac{dt}{dr}=\pm\sqrt{-\frac{g_{rr}}{g_{tt}}} \; .
\end{equation}
Their usage lies in the fact that they give the slope of the light-cones.
One of the characteristics of an event horizon is that the light-cones
fold up. Thus we demand that
\begin{equation}
-\frac{g_{rr}(r_H)}{g_{tt}(r_H)} \rightarrow \infty
\label{FC}
\end{equation}
at a horizon $r=r_H$. A further useful characterization of an event
horizon is its infinite redshift from an exterior observer's point of
view. Energies $E_1$ and $E_2$ belonging to the same physical process
but measured at radial positions $r_1$ and $r_2$ will differ by an
amount
\begin{equation}
\frac{E_1}{E_2}=\sqrt{\frac{g_{tt}(r_2)}{g_{tt}(r_1)}} \; .
\end{equation}
If we place $r_2=r_H$ at an event horizon and $r_1>r_H$
outside, an infinite redshift at the horizon implies that
\begin{equation}
g_{tt}(r_H)\rightarrow 0 \; .
\label{SC}
\end{equation}
This will serve as our second criterion. Let us now see which solutions
satisfy these two criteria and thus qualify as representing the exterior
stationary geometry of a horizon-located $D3$-$\Dta$ brane configuration.

The first case ($\Lambda=\Psi=0$) which gave the flat Minkowski spacetime
clearly has no horizon and moreover does not describe a localized
gravitational source as its ADM-mass vanishes. Thus we can rule it out.

The second case ($\Lambda\ne 0,\Psi=0$) of the embedded Schwarzschild
solution possesses a well-known event horizon at $r=r_0$ and thus
fulfills the above criteria.

The third case ($\Lambda=0,\Psi\ne 0$) gave no solutions.

In the fourth case ($\Lambda\ne 0,\Psi\ne 0,\Psi=c\Lambda$) the first
criterion (\ref{FC}) can only be satisfied if either
$K(r)\rightarrow\infty$ and $b>0$ or $K(r)\rightarrow\sqrt{2a^2-b}$. For
the first choice one learns from (\ref{Sol1}) that $r\rightarrow r_3$ and
thereby $g_{tt}$ does not vanish but approaches a positive value in
contradiction to (\ref{SC}). For the latter choice (\ref{Sol1}) says that
this is only possible at $r\rightarrow 0$. But to get a regular horizon
with finite area the brane set-up should be placed at some finite value
of $r_H>0$. Therefore this class of solutions has to be dismissed.

For the fifth solution ($\Psi=-\frac{1}{4}\Lambda$) $g_{tt}$ can only
vanish at $r\rightarrow 0$ which eliminates this solution as a candidate
for the same reason as for the second choice in the previous case.

As to the sixth solution ($\Psi=-\frac{2}{3}\Lambda$) the first criterion
(\ref{FC}) demands that $1\pm\frac{r_6}{r}\rightarrow 0$ while the second
criterion (\ref{SC}) would demand that
$1\pm\frac{r_6}{r}\rightarrow\infty$. This is impossible to fulfill
simultaneously which leads to the rejection also of this solution.

Since the Euclidean brane-antibrane pairs must give rise to some vacuum
solution we can now conclude that this will be the embedded D=4
Schwarzschild solution with the $D3-\Dta$ doublet placed at its horizon.
As has been stressed already in the introduction the main importance of
this identification lies in the fact that now we can straightforwardly
follow the general framework presented in \cite{K1} to derive the
BH-entropy and its corrections for the D=4 Schwarzschild solution by
counting chain-states on the worldvolume of the $D3$-$\Dta$ doublet
configuration.

\section{The Tachyon and Hawking-Radiation}
Let us finally speculate about the origin of the black hole's Hawking
radiation. It is well known that classically a black hole does not
radiate - it does so only when quantum effects are taken into account
\cite{Haw}. Let us try to understand this from our picture of the black
hole in terms of the Euclidean brane-antibrane pairs.

Coinciding brane-antibrane pairs suffer from an instability related
to the open string tachyon on their joint worldvolume. The tachyon
possesses a Mexican-hat like potential - be it quartic or more
complicated involving the error function \cite{TP}. Thus tiny
perturbations, e.g.~quantum fluctuations would initiate a rolling of the
tachyon down the hill if placed initially on top of the hill to describe
the brane-antibrane pair. According to Sen's conjecture \cite{Sen} the
brane-antibrane pair will have annihilated itself and produced just the
closed string vacuum by the time the tachyon reaches its potential
minimum by virtue of the equality
\begin{equation}
V(T_{min})=-2\frac{T_{Dp}}{g_s}
\label{SenConj}
\end{equation}
where $V(T_{min})$ is the negative value of the tachyon potential at its
minimum. In this tachyon-condensation process the relative U(1)
gauge-group under which the tachyon is charged gets spontaneously broken
by a Higgs-mechanism while the fate of the overall U(1) under which the
tachyon is neutral poses a problem \cite{Sred}. An interesting proposal
for its resolution is the idea that the overall U(1) gets confined
\cite{Yi}, leaving behind confined electric flux tubes which can be
identified with the fundamental closed strings \cite{FT} (see however
\cite{Shen} for criticism of this proposal).

Since the system starts with positive energy $2T_{Dp}/g_s$ (the tachyon
potential is zero on top of the hill) and ends up with zero energy due to
(\ref{SenConj}), it is commonly believed that the surplus of energy will
be radiated off into the bulk. For our concern the interesting aspect
lies in the fact that {\em classically} no such radiation will be
produced as shown recently in \cite{SenRad} but quantum mechanically
there is no obstruction and thus it should occur. This might be
understood from the fact that the process of radiation production has to
transform open strings on the brane-antibrane into closed strings which
can escape towards the bulk. However such an interaction arises only at
the {\em quantum level} (1-loop from the open string point of view) and
thus forbids a classical radiation into the bulk. Due to this coincidence
with the quantum appearance of Hawking radiation which is likewise
classically forbidden, it is natural to conjecture that Hawking radiation
in the black hole's brane-antibrane pair picture might be understood as
closed string modes send into the bulk when the tachyon rolls down
the hill.

It would be nice to make this more precise and also get a qualitative
understanding of the negative specific heat of the Schwarzschild black
hole from the brane description (see \cite{Dan} for an approach involving
a multitude of brane-antibrane pairs to explain the negative specific
heat for black 3-branes). It seems, however, that we have to acquire
first a better understanding of the dynamics of the tachyon condensation
process itself. For progress in this direction see
e.g.~\cite{SenRad},\cite{TD} or the interesting recent application of
brane-antibrane pairs to cosmology (see e.g.~\cite{COSM}). We hope to
report on progress on these issues and on a related embedding of de
Sitter spacetime in the future. Here, it would be interesting to find a
qualitatively different mechanism than suppression in warped geometries
\cite{KCC} to approach the problem of a finite but small vacuum energy.

\bigskip
\noindent {\large \bf Acknowledgements}\\[2ex]
The author wants to thank Allen Hatcher and Ashoke Sen for
correspondence. A.K.~associated to the Aristotle University of
Thessaloniki acknowledges financial support provided through the
European Community's Human Potential Program under contract
HPRN-CT-2000-00148 ``Physics Across the Present Energy Frontier'' and
support through the German-Greek bilateral program IKYDA-2001-22.

\newcommand{\atmp}[3]{{\sl Adv.Theor.Math.Phys.} {\bf C\,#1} (#2) #3}
\newcommand{\zpc}[3]{{\sl Z.Phys.} {\bf C\,#1} (#2) #3}
\newcommand{\npb}[3]{{\sl Nucl.Phys.} {\bf B\,#1} (#2) #3}
\newcommand{\npbps}[3]{{\sl Nucl.Phys.B(Proc.Suppl.)} {\bf #1} (#2) #3}
\newcommand{\plb}[3]{{\sl Phys.Lett.} {\bf B\,#1} (#2) #3}
\newcommand{\prd}[3]{{\sl Phys.Rev.} {\bf D\,#1} (#2) #3}
\newcommand{\prb}[3]{{\sl Phys.Rev.} {\bf B\,#1} (#2) #3}
\newcommand{\pr}[3]{{\sl Phys.Rev.} {\bf #1} (#2) #3}
\newcommand{\prl}[3]{{\sl Phys.Rev.Lett.} {\bf #1} (#2) #3}
\newcommand{\prsla}[3]{{\sl Proc.Roy.Soc.Lond.} {\bf A\,#1} (#2) #3}
\newcommand{\jhep}[3]{{\sl JHEP} {\bf #1} (#2) #3}
\newcommand{\cqg}[3]{{\sl Class.Quant.Grav.} {\bf #1} (#2) #3}
\newcommand{\prep}[3]{{\sl Phys.Rep.} {\bf #1} (#2) #3}
\newcommand{\fp}[3]{{\sl Fortschr.Phys.} {\bf #1} (#2) #3}
\newcommand{\nc}[3]{{\sl Nuovo Cimento} {\bf #1} (#2) #3}
\newcommand{\nca}[3]{{\sl Nuovo Cimento} {\bf A\,#1} (#2) #3}
\newcommand{\lnc}[3]{{\sl Lett.~Nuovo Cimento} {\bf #1} (#2) #3}
\newcommand{\ijmpa}[3]{{\sl Int.J.Mod.Phys.} {\bf A\,#1} (#2) #3}
\newcommand{\rmp}[3]{{\sl Rev. Mod. Phys.} {\bf #1} (#2) #3}
\newcommand{\ptp}[3]{{\sl Prog.Theor.Phys.} {\bf #1} (#2) #3}
\newcommand{\sjnp}[3]{{\sl Sov.J.Nucl.Phys.} {\bf #1} (#2) #3}
\newcommand{\sjpn}[3]{{\sl Sov.J.Particles\& Nuclei} {\bf #1} (#2) #3}
\newcommand{\splir}[3]{{\sl Sov.Phys.Leb.Inst.Rep.} {\bf #1} (#2) #3}
\newcommand{\tmf}[3]{{\sl Teor.Mat.Fiz.} {\bf #1} (#2) #3}
\newcommand{\jcp}[3]{{\sl J.Comp.Phys.} {\bf #1} (#2) #3}
\newcommand{\cpc}[3]{{\sl Comp.Phys.Commun.} {\bf #1} (#2) #3}
\newcommand{\mpla}[3]{{\sl Mod.Phys.Lett.} {\bf A\,#1} (#2) #3}
\newcommand{\cmp}[3]{{\sl Comm.Math.Phys.} {\bf #1} (#2) #3}
\newcommand{\jmp}[3]{{\sl J.Math.Phys.} {\bf #1} (#2) #3}
\newcommand{\pa}[3]{{\sl Physica} {\bf A\,#1} (#2) #3}
\newcommand{\nim}[3]{{\sl Nucl.Instr.Meth.} {\bf #1} (#2) #3}
\newcommand{\el}[3]{{\sl Europhysics Letters} {\bf #1} (#2) #3}
\newcommand{\aop}[3]{{\sl Ann.~of Phys.} {\bf #1} (#2) #3}
\newcommand{\jetp}[3]{{\sl JETP} {\bf #1} (#2) #3}
\newcommand{\jetpl}[3]{{\sl JETP Lett.} {\bf #1} (#2) #3}
\newcommand{\acpp}[3]{{\sl Acta Physica Polonica} {\bf #1} (#2) #3}
\newcommand{\sci}[3]{{\sl Science} {\bf #1} (#2) #3}
\newcommand{\nat}[3]{{\sl Nature} {\bf #1} (#2) #3}
\newcommand{\pram}[3]{{\sl Pramana} {\bf #1} (#2) #3}
\newcommand{\hepph}[1]{{\sl hep--ph/}{#1}}
\newcommand{\desy}[1]{{\sl DESY-Report~}{#1}}

\bibliographystyle{plain}

\end{document}